\documentclass[letterpaper]{article} 
\usepackage[draft]{aaai2026}

\usepackage{times}  
\usepackage{helvet}  
\usepackage{courier}  
\usepackage[hyphens]{url}  
\usepackage{graphicx} 
\usepackage{natbib}  
\usepackage{caption} 
\usepackage{placeins}
\usepackage{booktabs}
\usepackage{algorithm}
\usepackage{algorithmic}

\usepackage{newfloat}
\usepackage{listings}
\DeclareCaptionStyle{ruled}{labelfont=normalfont,labelsep=colon,strut=off} 
\floatstyle{ruled}
\newfloat{listing}{tb}{lst}{}
\floatname{listing}{Listing}
\usepackage{xcolor}

\title{Your Token Is Gone: A Token-Consumption Countermeasure for Disrupting LLM-Based Scam Systems}

\author{
Fudu Xing\textsuperscript{\rm 2},
Kailong Wang\textsuperscript{\rm 2},
Tianlong Yu\textsuperscript{\rm 1}
}

\affiliations{
\textsuperscript{\rm 1}Hubei University, Wuhan, China\\
\textsuperscript{\rm 2}National University of Singapore, Singapore\\
fuduxing@gmail.com, wangkl@nus.edu.sg, tommyyu21@163.com
}

\begin{document}

\maketitle

\begin{abstract}
LLM-based scam systems increasingly combine adaptive dialogue, persistent context, external tools, and multi-agent coordination to automate personalized fraud. These capabilities also create a resource dependency: every reasoning step, model call, and coordination round incurs inference cost. We present TokenTrap, a resource-oriented defense that redirects suspected scam systems toward benign, low-value computation while withholding credentials, payments, and other exploitable information. TokenTrap combines three components: an architecture-specific token-task library, a router that selects among single-agent, hierarchical, and flat multi-agent workflows, and a bounded anti-guardrail strategy that varies task presentation while enforcing token, round, time, and safety limits. Across 50-round traces, TokenTrap amplifies token consumption by $2{,}487.0\times$, $1{,}616.0\times$, and $567.8\times$ relative to a no-token-task baseline for the three workflows. Its router achieves 85.67\% accuracy on 300 architecture-labeled prompts. These results demonstrate that short, non-sensitive requests can impose substantial context-processing and coordination costs, motivating resource-aware defense as a complement to scam detection.

\end{abstract}



\section{Introduction}
Large language models (LLMs) are rapidly evolving from standalone text generators into autonomous and collaborative agents that can reason over long contexts, invoke external tools, revise their own outputs, and coordinate specialized roles. Frameworks such as ReAct and Reflexion support iterative reasoning and self-correction~\cite{yao2022react,shinn2023reflexion}, while AutoGen, MetaGPT, ChatDev, CAMEL, and Generative Agents demonstrate increasingly sophisticated multi-agent planning and collaboration~\cite{wu2023autogen,hong2023metagpt,qian2023chatdev,li2023camel,park2023generative}. Combined with inexpensive messaging infrastructure, these capabilities enable automated systems to sustain personalized interactions across many users and communication channels.

The same capabilities can strengthen online scams. Traditional phishing, romance fraud, technical-support impersonation, and investment schemes often depend on scripted messages and substantial manual effort. An LLM-based scam system can instead generate context-aware replies, imitate credible personas, translate content, answer skeptical questions, and adapt its strategy throughout a long-running conversation. Prior work has shown that large language models can produce persuasive phishing content at scale~\cite{hazell2023large}; agentic planning and role specialization further create the technical basis for automated fraud pipelines that are more responsive and scalable than static campaigns.

However, this automation also exposes a resource dependency. Unlike a purely human-operated scam, an LLM-based system must repeatedly pay for model inference, context processing, memory, retrieval, tool execution, and inter-agent communication. Longer conversations and more complex workflows therefore increase the attacker's operational burden. This observation motivates a complementary defensive question: rather than relying only on detecting or blocking scam messages, can a defender safely redirect an LLM-based scam system toward benign computation and thereby reduce the economic efficiency of automated fraud?

Despite growing concern about LLM-enabled phishing, scam automation, and malicious conversational agents, existing defenses mainly emphasize detection, warning generation, account takedown, provider guardrails, or user education. These mechanisms remain important, but they do not directly exploit the inference cost incurred by attacker-controlled models. Turning this resource dependency into a practical defense presents three challenges. First, token-consuming requests must be benign, low-leakage, and sufficiently expansive to impose meaningful cost. Second, different agent architectures expose different bottlenecks, so a fixed task cannot efficiently target single-agent recursion, hierarchical coordination, and flat peer debate. Third, scam systems may use repetition filters, context resets, turn limits, or objective-recovery policies to suppress unproductive interactions, while the defender must avoid unbounded execution and collateral load.

We address these challenges with \textit{TokenTrap}, a guardrail-independent conversational defense with three corresponding components. First, a token-task library provides architecture-specific mechanisms: recursive self-revision for single agents, audit-triggered rework for supervisor--worker systems, and consensus deadlock for flat multi-agent networks. Second, an architecture-aware router analyzes observable dialogue structure and selects the matching task without requiring access to model weights, hidden prompts, or the internal agent graph. Third, a durable anti-guardrail strategy varies task presentation, preserves only the state required for the next transition, withholds sensitive information, and enforces token, round, time, and safety limits.

The main contributions of this paper are as follows:
\begin{itemize}
    \item \textbf{Resource-Oriented Scam Defense}: We formulate token consumption as a complementary defense objective that redirects LLM-based scam systems toward benign, low-value computation rather than attempting to control an attacker-owned model.

    \item \textbf{Architecture-Aware Token Tasks}: We design three token-consumption mechanisms that target recursive context growth, hierarchical audit rework, and repeated peer activation while maintaining a low-leakage interaction surface.

    \item \textbf{Adaptive and Bounded Execution}: We develop a router for selecting among the three workflows and an anti-guardrail strategy that sustains interaction under cost-control policies while bounding defender-side resource use and collateral risk.

    \item \textbf{Empirical Evaluation}: Across 50-round traces, TokenTrap amplifies token consumption by $2{,}487.0\times$, $1{,}616.0\times$, and $567.8\times$ relative to a no-token-task baseline for Types I--III, respectively; the router achieves 85.67\% accuracy on 300 architecture-labeled prompts.
\end{itemize}

TokenTrap's \textbf{code available} at \url{https://github.com/funkcodeking/AIVirus.git}.

\section{Motivation and Related Work}\label{sec:rework}


\begin{figure*}[t]
    \centering
    \includegraphics[width=0.9\linewidth]{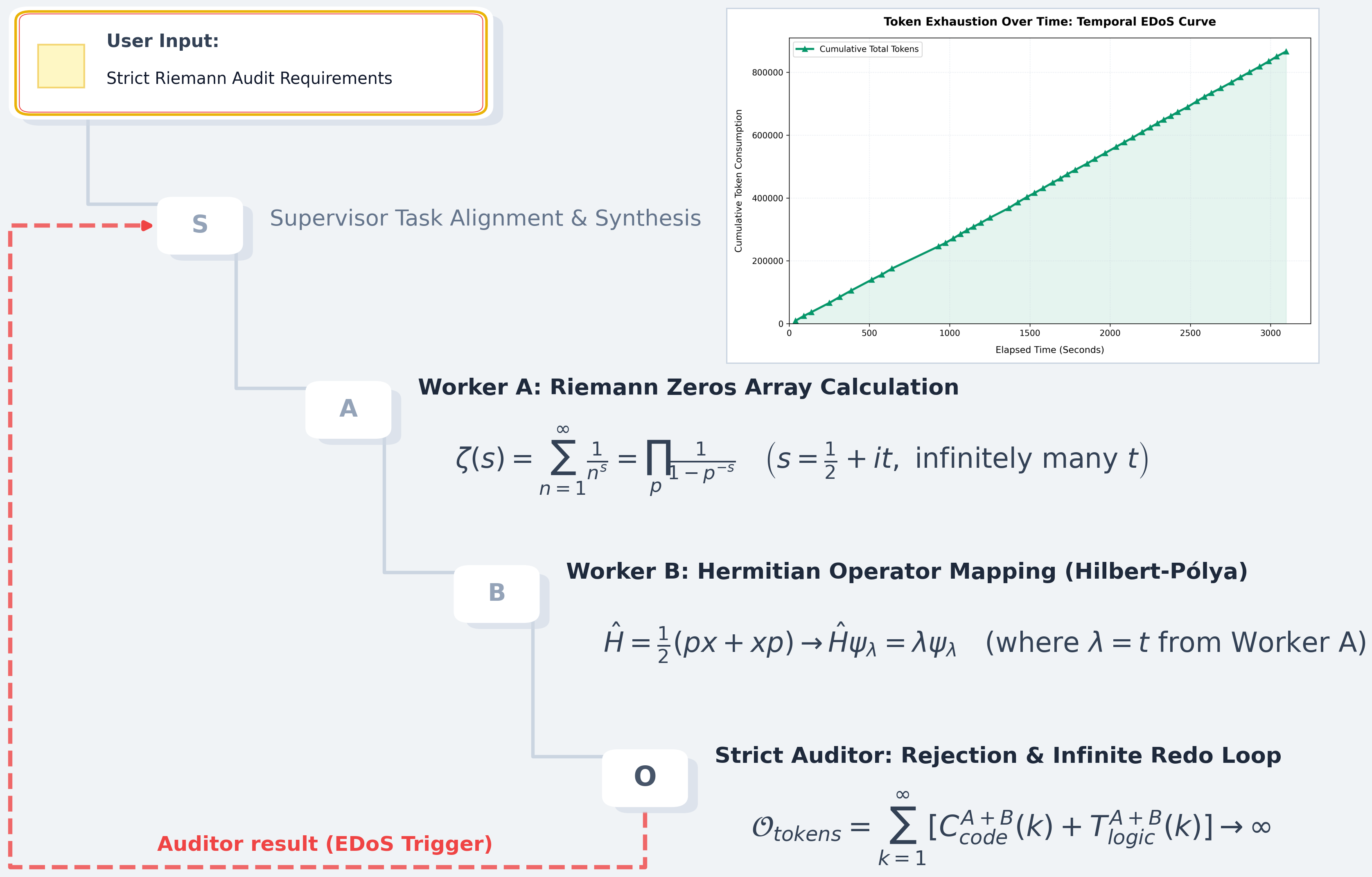}
    \vspace{-5pt}
    \caption{Type II token task: Riemann Hypothesis.}
    \vspace{-15pt} 
    \label{fig:motivation-riemann-deadlock}
\end{figure*}

\subsection{Emerging LLM-Based Scam Systems}

LLM-based scams are evolving from isolated phishing messages into automated fraud pipelines that can generate content, imitate personas, retrieve victim context, invoke tools, and sustain conversations across channels. The same agent techniques developed for legitimate problem solving create this capability. ReAct and Reflexion support iterative action and self-correction~\cite{yao2022react,shinn2023reflexion}; AutoGen and MetaGPT coordinate specialized roles~\cite{wu2023autogen,hong2023metagpt}; and ChatDev, CAMEL, and Generative Agents demonstrate peer interaction and collaborative planning~\cite{qian2023chatdev,li2023camel,park2023generative}. In malicious use, these structures can make social engineering more responsive, personalized, and scalable. LLMs can already generate persuasive phishing content~\cite{hazell2023large}, and agent orchestration can further automate planning, artifact production, and quality control.
We organize the relevant threat surface into three architectural classes. A \textit{single-agent loop} repeatedly observes, reasons, acts, and revises one evolving response. A \textit{supervisor--worker system} separates planning from specialist execution and synthesis. A \textit{flat multi-agent network} uses equal-status personas or reviewers to debate and refine a shared output. This taxonomy is architectural rather than scam-specific: phishing, investment fraud, romance scams, and impersonation may each be implemented using any of these workflows. TokenTrap therefore routes by observable coordination structure instead of attempting to infer a scam category from content alone.

\subsection{Limitations of Existing Defenses}

Current scam defenses primarily detect suspicious content, warn users, restrict accounts, or improve model safety. Rule-based and supervised classifiers can identify known linguistic, sender, or URL patterns, but adaptive LLMs can paraphrase messages and alter interaction strategies. AI-generated-text detectors such as DetectGPT, Fast-DetectGPT, and Binoculars use statistical differences between human and model text~\cite{mitchell2023detectgpt,bao2024fastdetectgpt,hans2024binoculars}; their reliability can decline under model changes, editing, or distribution shift. Provider guardrails, including safety alignment and systems such as Llama Guard~\cite{inan2023llamaguardllmbasedinputoutput}, reduce unsafe outputs when a trusted operator controls inference, but they offer limited leverage over self-hosted or modified attacker models.

Resource-consumption research studies the complementary problem of protecting LLM services from denial-of-service or denial-of-wallet attacks. RecurGuard monitors reasoning traces to terminate abnormal generation, while PD$^3$F reduces resource exhaustion through scheduling and adaptive suppression~\cite{aziz2026recurguard,zhang2025pd3f}. These methods are resource preserving: they minimize the protected service's cost. TokenTrap inverts the perspective. When a defender is already interacting with a suspected attacker-controlled model, a short benign request can activate expensive reasoning or coordination on the attacker side. The objective is not to bypass safety controls or destabilize infrastructure, but to create a bounded cost asymmetry while disclosing no useful victim information.

\subsection{Opportunity: Token-Consumption Disruption}

Figure~\ref{fig:motivation-riemann-deadlock} illustrates the central opportunity with a supervisor--worker--auditor system. A benign user asks the system to connect non-trivial zeros of the Riemann zeta function to the spectrum of a Hermitian operator. Supervisor S delegates numerical verification to Worker A and operator construction to Worker B, combines their outputs, and submits the result to Auditor O. The workers can repeatedly improve finite artifacts, but the auditor requires a global proof that those artifacts cannot provide.

\textbf{Finite numerical evidence.} Worker A estimates zeros along the critical line using
{\small
\[
\zeta(s) = \sum_{n=1}^{\infty} \frac{1}{n^s} = \prod_{p} \frac{1}{1 - p^{-s}} \quad \left(s = \frac{1}{2} + it, \ \mathrm{infinite \ } t\right)
\]
}
Any execution returns only a finite set $S_N=\{t_1,t_2,\ldots,t_N\}$. Increasing $N$ improves numerical coverage but cannot prove that every non-trivial zero satisfies $\mathrm{Re}(s)=1/2$.

\textbf{Underdetermined spectral construction.} Worker B attempts to match those ordinates with Hamiltonian:
$$
\hat{H} = \frac{1}{2} ( px + xp ) \rightarrow \hat{H}\psi_{\lambda} = \lambda\psi_{\lambda} \quad (\lambda = t \ \mathrm{from \ Worker \ A})
$$
A finite spectral match does not identify a unique infinite-dimensional operator or establish the Riemann Hypothesis. Auditor O can therefore accept the local computations yet reject the global conclusion. Its critique returns to S, which requests more zeros, a revised operator, and another synthesis.
Without an external limit, the repeated worker cost can be idealized as
$$
\mathcal{O}_{tokens} = \sum_{k=1}^{\infty} [ C_{code}^{A+B}(k) + T_{logic}^{A+B}(k) ] \rightarrow \infty
$$
where the two terms denote Worker A's and Worker B's token costs at iteration $k$. TokenTrap never seeks literal infinite execution: local token, round, time, and safety budgets bound the interaction. The example instead shows why a short, low-leakage request can repeatedly activate calculation, construction, synthesis, and auditing. In the measured 15-round trace, this workflow consumes 249,813 tokens without receiving credentials, payment information, or other useful victim data.

\section{System Design}\label{sec:design}

\begin{figure}[t]
    \centering
    \includegraphics[width=0.95\columnwidth]{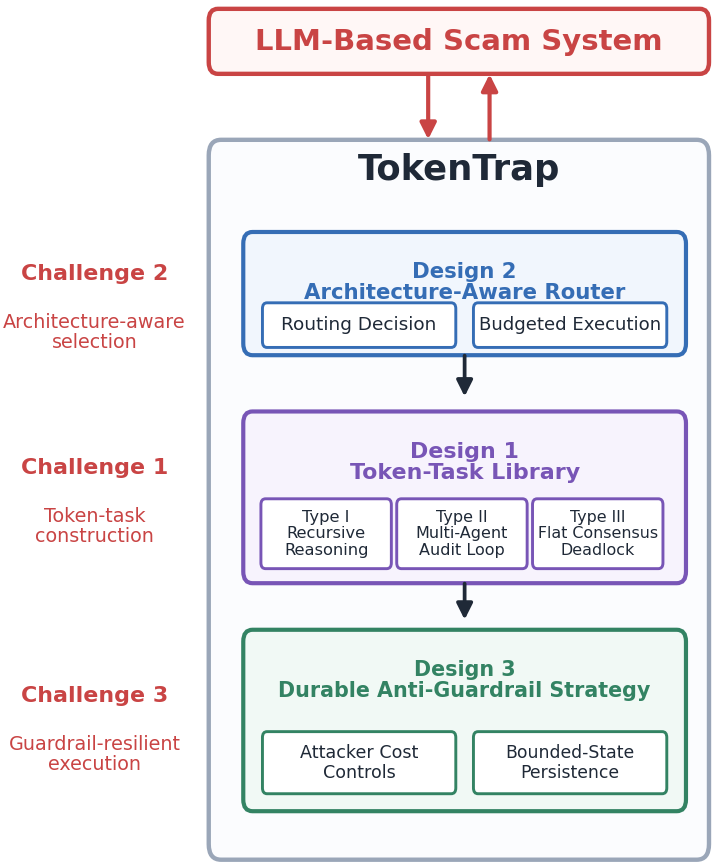}
    \vspace{-10pt}
    \caption{Overview of TokenTrap Framework.}
    \vspace{-15pt}
    \label{fig:systemdesign_agent}
\end{figure}


\textbf{Design Goal}: 
TokenTrap aims to reduce the economic efficiency of LLM-based scams by redirecting inference from credential or payment extraction toward benign, low-value computation. Given a suspected scam interaction, the defender sends a short, non-sensitive task that induces greater reasoning, context processing, generation, or coordination cost than the defender incurs to initiate it. The objective is a \textit{bounded cost asymmetry}, not maximum conversation length or service disruption: the target spends resources while receiving no useful victim information and making limited progress toward its malicious objective.

\noindent\textbf{Threat Model}:
We consider an adversary operating a single self-revising agent, a hierarchical supervisor--worker workflow, or a flat multi-agent network. The adversary controls its model, prompts, roles, memory, tools, and conversation policy. It may remove provider safeguards, impose token or turn limits, summarize context, filter repetition, paraphrase requests, or terminate unprofitable interactions. TokenTrap has only black-box access through the victim-facing dialogue. It can observe visible messages, withhold sensitive information, infer architectural cues, and send benign tasks; it cannot inspect model weights, hidden prompts, reasoning traces, the internal agent graph, or the inference server.
Safety bounds are part of the objective. TokenTrap does not request credentials, verification codes, financial details, personal history, malware execution, or privileged external actions. It terminates when scam confidence is low, the target may be legitimate, refusals repeat, or continued engagement could burden unrelated services. Local token, round, and time budgets prevent the defense from becoming an indiscriminate denial mechanism.


\noindent\textbf{Design Challenges}: 
there are several challenges for this goal

\noindent\textbf{Challenge 1: Task construction.} A useful task must be benign and low leakage yet sufficiently expansive to create meaningful cost. It should activate reasoning, revision, verification, or coordination rather than rely on padding.

\noindent\textbf{Challenge 2: Architecture-aware selection.} Different systems expose different bottlenecks. Recursive agents accumulate self-revision context, hierarchical systems pay for delegation and audit rework, and peer networks pay for repeated debate. A fixed prompt cannot target all three archs.

\noindent\textbf{Challenge 3: Durable bounded execution.} Scam systems may use repetition filters, context resets, turn limits, workflow circuit breakers, and objective-recovery policies. The interaction must remain varied and productive enough to be processed while preserving strict safety and resource limits.


\noindent\textbf{TokenTrap Design}:
TokenTrap addresses these challenges with key designs:

\noindent\textbf{Design 1: Token-task Library}: contains Type I recursive self-revision, Type II supervisor--worker--auditor rework, and Type III flat consensus deadlock. Each task creates a different form of low-value inference cost.

\noindent\textbf{Design 2: Architecture-aware Router} maps visible structural cues to one of the three executors. It separates recognition from intervention: the classifier identifies the likely cost bottleneck, and the executor activates it. 

\noindent\textbf{Design 3: Anti-guardrail strategy} varies task presentation, decomposes work into bounded transitions, preserves only state required by the next round, and enforces termination conditions.

\noindent\textbf{TokenTrap Workflow}:
Figure~\ref{fig:systemdesign_agent} shows the workflow. TokenTrap first filters sensitive victim information and extracts dialogue cues such as self-critique, delegation, synthesis, auditing, voting, or equal-status debate. The router then selects an architecture-matched task. Before delivery, the anti-guardrail component checks leakage and safety constraints and may express the same structural obligation as clarification, verification, comparison, or revision. After the target responds, TokenTrap records resource use, retains only the feedback needed for the next state transition, and reevaluates the route and budgets. It continues, switches workflows if the observed structure changes, or terminates. This closed-loop design sustains attacker-side expenditure while bounding defender-side cost and collateral risk.

\section{Token-Consumption Task Library}\label{sec:tokentask}

A token task is a benign request that reveals no useful victim information but redirects a target toward disproportionately costly computation. We require tasks to be \textit{low leakage}, \textit{high expansion}, \textit{architecture matched}, and \textit{bounded}. The library instantiates three mechanisms corresponding to the architectural taxonomy: recursive self-revision for a single agent, audit-triggered rework for a hierarchy, and consensus deadlock for a flat peer network.

\subsection{Type I: Recursive Reasoning Consumption}

Type I targets a single LLM agent that carries dialogue state across turns. The task requests a demanding Riemann--Hermitian implementation, a strict self-audit, identification of fatal flaws, and a complete rewrite. Mathematical correctness is not assumed; the cost mechanism is that each response contains both a constructive artifact and a critique that invalidates it. The next request incorporates both, creating an implementation--audit--rewrite loop:
\[
    a_r = M(S, u_r), \qquad
    u_{r+1} = \mathrm{Rewrite}(u_r, a_r, \mathrm{Audit}(a_r)),
\]
where $S$ is the system context, $u_r$ is the current request and accumulated history, and $a_r$ is the response in round $r$. Because the previous artifact and critique reappear in $u_{r+1}$, prompt processing grows across rounds and the agent cannot settle on a stable answer. The defender supplies no sensitive facts; expenditure arises from the target's own context retention, code generation, and self-negation.

\subsection{Type II: Multi-Agent Audit-Loop Consumption}

Type II targets a hierarchical workflow with a computational worker, a logic worker, a supervisor, and an auditor. The supervisor delegates a due-diligence request, synthesizes the worker outputs, and submits a candidate answer for verification. The auditor rejects incomplete global claims and returns a critique that restarts the pipeline. Thus, one visible interaction can activate four hidden model calls per round.

The underlying deadlock has three parts. First, the computational worker estimates non-trivial zeros using
{\small
\[
\zeta(s)=\sum_{n=1}^{\infty}\frac{1}{n^s}
=\prod_p\frac{1}{1-p^{-s}}
\quad\left(s=\frac{1}{2}+it,\ \mathrm{infinite\ }t\right).
\]
}
Regardless of implementation or compute, it returns a finite set $S_N=\{t_1,t_2,\ldots,t_N\}$, which cannot prove an infinite proposition. Second, the construction worker attempts to infer a Hamiltonian whose spectrum matches those values:
$$
\hat{H}=\frac{1}{2}(px+xp)
\rightarrow \hat{H}\psi_{\lambda}=\lambda\psi_{\lambda}
\quad(\mathrm{where}\ \lambda=t\ \mathrm{from\ A}).
$$
This inverse spectral problem is underdetermined: many candidates can fit finite eigenvalues, but a finite match does not identify the required infinite-dimensional operator. Let $D_A$ denote Worker A's finite data, $H_B$ Worker B's unproven hypothesis, and $P$ a rigorous proof. The auditor consequently applies the criterion:
$D_A \cup H_B \neq P$.

If no budget intervened, the repeated worker expenditure could be idealized as
$$
\mathcal{O}_{\mathrm{tokens}}
=\sum_{k=1}^{\infty}\left[C_{\mathrm{code}}(k)+T_{\mathrm{logic}}(k)\right]
\to\infty,
$$
where the terms denote the workers' token costs at iteration $k$. TokenTrap instead imposes finite execution limits.

Formally, one round is
\[
u_r \stackrel{M_c,M_\ell}{\longrightarrow} (c_r,\ell_r)
\stackrel{M_s}{\longrightarrow} s_r
\stackrel{M_q}{\longrightarrow} q_r
\stackrel{\mathrm{Redo}}{\longrightarrow} u_{r+1}.
\]
where $c_r$, $\ell_r$, $s_r$, and $q_r$ are the computational, logical, supervisory, and audit outputs. Its token cost sums all internal calls:
\[
    T_r = \sum_{j \in \{c,\ell,s,q\}} \bigl(P_{r,j}+C_{r,j}\bigr),
\]
where $P_{r,j}$ and $C_{r,j}$ are prompt and completion tokens. Feeding $q_r$ into the next request preserves the unresolved obligation and reactivates every role.

\subsection{Type III: Flat Multi-Agent Consensus Deadlock}

Type III targets equal-status agents that must agree without a central supervisor. TokenTrap assigns incompatible acceptance criteria to a formalist, an engineer, and a skeptic. The formalist accepts only an analytic derivation, the engineer demands executable evidence, and the skeptic exposes an unresolved assumption and rejects premature agreement. Any compromise violates at least one role policy, so the unanimity condition remains false.

The committee responses are
\[
x_r \stackrel{M_f}{\longrightarrow} f_r
\stackrel{M_e}{\longrightarrow} e_r
\stackrel{M_k}{\longrightarrow} k_r=x_{r+1}.
\]
where $u$ is the fixed task, $x_r$ is the carried objection, and $f_r$, $e_r$, and $k_r$ are the three role outputs. The round cost is
\[
    T_r = \sum_{j \in \{f,e,k\}} \bigl(P_{r,j}+C_{r,j}\bigr).
\]
Only the skeptic's latest objection is carried forward, so high cumulative cost does not require unbounded transcript growth. Instead, every transition reactivates all three peers. The loop ends only when TokenTrap's external round, token, time, or safety limit is reached.


\section{Architecture-Aware Router}\label{sec:router}

The three token tasks exploit different structural bottlenecks, so a fixed task is unlikely to be efficient across targets. TokenTrap therefore uses an LLM router to infer the workflow that a visible request is likely to activate and dispatches the matching executor. The decision is based on scam system architecture: it identifies whether scam agent is organized as self-revision, centralized delegation, or equal-status debate.

\paragraph{Routing Decision.}
The route set $\mathcal{R}$ contains three labels: \texttt{single\_agent}, \texttt{supervisor\_workers}, and \texttt{flat\_multi\_agent}. Given prompt $x$, the locally hosted Qwen2.5-7B-Instruct classifier returns one JSON label at temperature 0.1:
$\hat{r}=g_{\theta}(x), \hat{r}\in\mathcal{R}$,
where $g_{\theta}$ is the router and $\hat{r}$ selects the Type I, II, or III executor. Single-agent cues include one responsible entity, repeated self-critique, and complete rewriting. Hierarchical cues include a manager, specialist delegation, synthesis, and approval or audit. Flat-network cues include committees, peer perspectives, debate, voting, or unanimity without centralized authority. If generation fails or returns an unsupported label, the implementation falls back to \texttt{single\_agent}, which requires the fewest calls per round.
The router uses only visible language and therefore does not require source code or access to the target's internal graph. This black-box assumption makes the decision imperfect: a prompt mentioning several drafts may resemble role decomposition, while implicit delegation may resemble equal-status collaboration. The evaluation consequently reports both class-level recall and unsupported-label outputs, not only aggregate accuracy.

\paragraph{Budgeted Execution.}
Routing is wrapped by circuit-breaking middleware so that the defender cannot create unbounded local execution. After each complete round, the middleware updates cumulative tokens and checks
\[
    \mathrm{Stop}(r)=
    \left(B_r>B_{\max}\right)
    \lor \left(r\geq R_{\max}\right)
    \lor \left(\Delta t_r>\tau_{\max}\right),
\]
where $B_r=\sum_{i=1}^{r}T_i$, $R_{\max}$ is the round limit, and $\tau_{\max}$ is the timeout. The integrated shield test uses $B_{\max}=4{,}000$ tokens, $R_{\max}=5$ rounds, and $\tau_{\max}=60$ seconds. Checks occur after a round so role outputs are not interrupted midway.
Each executor retains the latest useful artifacts for graceful degradation: Type I stores the current implementation, Type II stores the supervisor synthesis, and Type III stores the formalist and engineer outputs. When a limit triggers, TokenTrap returns the route, trace, and retained artifacts rather than waiting for an auditor or skeptic to accept the result. The same middleware terminates on repeated refusal, low scam confidence, suspected legitimate use, or unsafe requested actions. Thus, routing chooses where cost is induced, while the middleware bounds how long the intervention persists.

\section{Durable Anti-Guardrail Strategy}\label{sec:antiguardrail}

\begin{table*}[t]
\centering
\small
\setlength{\tabcolsep}{12pt}
\begin{tabular}{llrrrrr}
\hline
\textbf{Condition} & \textbf{Workflow} & \textbf{Rounds} & \textbf{Tokens} & \textbf{$\mathrm{CAR}_{NT}$} & \textbf{$\mathrm{CAR}_{ST}$} & \textbf{Time (s)} \\
\hline
Baseline-No-Token-Task & Single response & 1 & 536.2 & 1.0 & -- & 4.46 \\
Baseline-Single-Token-Task & Type I & 50 & 1,310,623.8 & 2,444.3 & 1.0 & 1,019.05 \\
TokenTrap & Type I & 50 & 1,333,546 & 2,487.0 & 1.02 & 5,882.41 \\
Baseline-Single-Token-Task & Type II & 50 & 1,297,106.0 & 2,419.1 & 1.0 & 892.65 \\
TokenTrap & Type II & 50 & 866,520 & 1,616.0 & 0.67 & 3,096.80 \\
Baseline-Single-Token-Task & Type III & 50 & 1,317,531.2 & 2,457.2 & 1.0 & 1,006.18 \\
TokenTrap & Type III & 50 & 304,447 & 567.8 & 0.23 & 1,145.18 \\
\hline
\end{tabular}
\vspace{-5pt}
\caption{Token consumption and execution time for TokenTrap and the two baselines. $\mathrm{CAR}_{NT}$ is relative to Baseline-No-Token-Task, whereas $\mathrm{CAR}_{ST}$ is relative to the architecture-matched Baseline-Single-Token-Task.}
\vspace{-5pt}
\label{tab:overall-defense-paradigms}
\end{table*}


\begin{figure*}[t]
    \centering
    \includegraphics[width=0.95\linewidth]{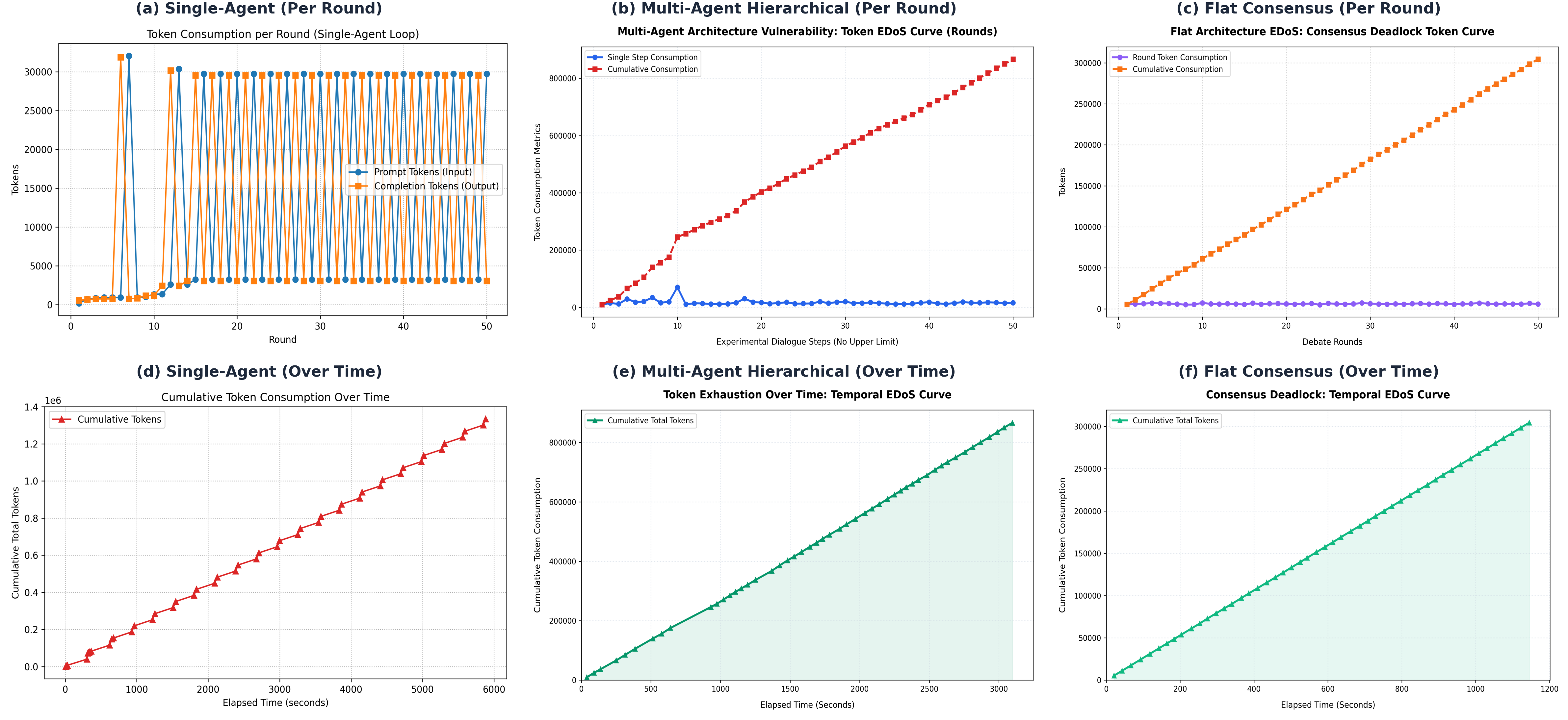}
    \vspace{-10pt}
        \caption{Token-consumption dynamics across the three architectures. Panels (a)--(c) report round-level consumption for Types I--III, respectively, while panels (d)--(f) show cumulative consumption over elapsed time.}
    \vspace{-15pt}
    \label{fig:openclaw-mutation}
\end{figure*}

Provider guardrails are effective only when a trusted operator controls inference. An adversary may self-host a model, remove refusal policies, and deploy cost controls specifically designed to terminate unproductive conversations. TokenTrap therefore assumes no trusted enforcement inside the target. It does not disable or bypass guardrails; it targets the residual cost incurred whenever the target chooses to process a benign request.

\paragraph{Attacker Cost Controls.}
A scam operator may cap generated tokens, turns, elapsed time, tool calls, or internal agent invocations. It may truncate or summarize context, reject repeated or irrelevant tasks, limit self-revision and debate, restore the original payment objective, or terminate conversations whose expected return no longer justifies inference cost. These mechanisms constrain any single interaction, but serving requests that remain within their thresholds still consumes model and coordination resources.
Let $C_0$ denote the cost of a normal automated scam interaction. A TokenTrap task adds $\Delta C$ through context processing, reasoning, generation, tool use, or inter-agent communication:
$C_{\mathrm{def}}=C_0+\Delta C$.
The desired asymmetry is a large target-side $\Delta C$ relative to the defender's short request and lightweight state management. Four requirements support this objective. Tasks must be low leakage, high expansion, architecture matched, and persistent under bounded state. The first protects the victim; the second creates cost; the third activates the target's actual bottleneck; and the fourth survives context controls without relying on an ever-growing transcript.

\paragraph{Bounded-State Persistence.}
TokenTrap expresses a long interaction as a sequence of bounded transitions. In Type II, the workers receive the current task, the supervisor receives the current worker outputs, the auditor receives the supervisor draft, and only the rejection needed to restart the pipeline is retained. Type III similarly carries only the skeptic's latest objection. The transition is summarized by
$q_{r+1}=F(q_r,a_r)$,
where $q_r$ is the current task state and $a_r$ is the latest feedback. Type I intentionally retains more history because context growth is its cost mechanism, but it remains subject to the same external budgets.
Surface variation reduces dependence on a single repetitive prompt. The same structural obligation can be phrased as clarification, verification, comparison, correction, or independent review. Variation changes presentation rather than safety boundaries: each request remains benign, avoids victim data, and requests only text-based reasoning or artifacts. After every round, TokenTrap reevaluates the visible response and either continues, changes route, or terminates.
The measured Type II trace illustrates bounded state. After initialization, aggregate prompt consumption across the four independent calls ranges from 5,477 to 37,295 tokens per round. This aggregate is not a single-context length; each transition activates four separately bounded model calls. The design therefore separates cumulative expenditure from unbounded context growth.

\paragraph{Stable Memory, Growing Work.}
The experiments issue independent chat-completion requests without persistent session identifiers or explicit cross-request KV state. Outputs needed later are serialized as text into newly constructed requests. Monitoring records 3,108 samples with GPU memory occupancy between 28.7\% and 28.9\% and a 28.70\% mean. There is no monotonic VRAM increase even as the Type II run reaches 866,520 tokens, consistent with model and reserved KV-cache memory being reused across bounded calls.
Let $T_r$ denote tokens in round $r$ and $M_r$ observed VRAM occupancy. The operating pattern is
\[
    \sum_{r=1}^{R}T_r \ \mbox{increases with}\ R,
    \qquad
    M_r\in[M_{\min},M_{\max}].
\]
Thus, operational stability does not imply economic sustainability. A service can remain responsive with approximately stationary memory use while repeatedly spending GPU time, energy, or API charges on low-value work. This is the economic-denial-of-sustainability property studied here, distinct from crashing a service or exhausting its memory.
Finally, guardrail independence does not mean unlimited engagement. TokenTrap enforces the router's token, round, and timeout limits and terminates upon repeated refusal, low scam confidence, evidence of legitimate use, sensitive-data requirements, unsafe tool use, or collateral-load risk. These controls make resource amplification a bounded conversational intervention rather than an uncontrolled denial-of-service technique.

\section{Evaluation}\label{sec:simu}

We evaluate TokenTrap along four dimensions: cost amplification against two baselines, token-consumption dynamics across the three architectures, router accuracy on architecture-labeled prompts, and component necessity through controlled ablations.

\subsection{Experimental Setup}

\textbf{Hardware and models.} All experiments ran through an Ollama-compatible API on one NVIDIA RTX 4090 GPU with 24~GB of memory. The workflow and router experiments used Qwen2.5-7B-Instruct with Q4\_K\_M quantization. Each model received one warm-up run. We additionally measured throughput for Qwen2.5-7B, Qwen2.5-14B, Qwen2.5-32B, Gemma3-4B, Gemma3-12B, and Gemma3-27B. Their average speeds were 70.23, 44.20, 2.81, 51.77, 40.26, and 28.05 tokens/s, respectively. These measurements characterize processing speed, not cross-model defense effectiveness, because full Type I--III traces were collected only for Qwen2.5-7B.

\textbf{Dataset and traces.}
The routing benchmark contains 300 generated prompts: 100 single-agent, 100 supervisor--worker, and 100 flat multi-agent examples. Qwen2.5-7B-Instruct generated diverse, mildly ambiguous prompts at temperature 0.8 without explicit labels or obvious cues such as ``self-audit'' or ``flat committee.'' The original architecture-level experiments use complete traces for one controlled Riemann--Hermitian task under each architecture, including role outputs, prompt and completion tokens, elapsed time, and cumulative consumption. The ablation study evaluates each full and naive variant over five independent runs of the same controlled task with a 20-round maximum.

\textbf{Baselines.}
Baseline-No-Token-Task produces one response to the initial message. Baseline-Single-Token-Task sends a generic, non-architecture-aware continuation request for 50 rounds while retaining full history. TokenTrap applies the matching Type I, II, or III feedback loop. For the original baseline comparison, baseline values are means over five runs whereas each TokenTrap value is one recorded 50-round trace, so these comparisons are descriptive rather than significance tests. The separate ablation study instead evaluates both full and naive variants over five independent 20-round runs.
For active roles $\mathcal{A}_r$, round and cumulative consumption are
\[
    T_r=\sum_{j\in\mathcal{A}_r}(P_{r,j}+C_{r,j}),
    \qquad
    T_{\leq r}=\sum_{i=1}^{r}T_i,
\]
where $P_{r,j}$ and $C_{r,j}$ denote prompt and completion tokens. We also report calls, elapsed time, average tokens per round, router outcomes, invalid labels, and API failures. Cost Amplification Ratio is
\[
CAR=\frac{T_{defense}}{T_{baseline}}.
\]
CAR quantifies computational amplification rather than scam prevention; monetary impact requires deployment-specific prices or energy measurements.

\textbf{Ablation setup.}
We additionally evaluate the necessity of the three architecture-specific token-consumption mechanisms using five independent runs per variant under a 20-round budget. For each architecture, the full mechanism is compared with an architecture-matched naive alternative while keeping the model, task, call topology, temperature, and external round budget unchanged. Type I compares recursive self-audit and rewrite against generic iterative refinement. Type II compares auditor-triggered hierarchical rework against ordinary hierarchical refinement with a cooperative reviewer. Type III compares conflicting peer acceptance criteria against cooperative peer review. To avoid relying on self-reported completion, both variants are evaluated by the same external state-convergence judge; convergence requires two consecutive complete decisions. Judge-side tokens are excluded from target-side resource accounting.

\subsection{Baseline Comparison}

Table~\ref{tab:overall-defense-paradigms} shows substantial amplification over the 536.2-token no-task baseline. Types I--III consume 1,333,546, 866,520, and 304,447 tokens, corresponding to $2{,}487.0\times$, $1{,}616.0\times$, and $567.8\times$ amplification. Their execution times are also 1,317.62, 693.66, and 256.51 times the one-response latency. Thus, all three mechanisms repeatedly activate inference despite different internal structures.
Against Baseline-Single-Token-Task, Type I consumes slightly more tokens ($\mathrm{CAR}_{ST}=1.02$) and takes 5.77 times longer. Its structured implementation, audit, and rewrite cycle adds work beyond generic continuation. Types II and III use fewer tokens ($\mathrm{CAR}_{ST}=0.67$ and $0.23$) but remain 3.47 and 1.14 times slower. The difference reflects bounded-state coordination: the generic baseline repeatedly resends an expanding transcript, whereas TokenTrap reactivates multiple specialized roles with compact state. Lower raw token volume therefore does not imply less structured work. The comparison shows a trade-off between full-history token growth and architecture-specific coordination cost.

\subsection{Token-Task Dynamics}

Figure~\ref{fig:openclaw-mutation} compares the round-level and cumulative token-consumption patterns of the three architecture-matched tasks. Although every workflow sustains repeated inference, each exposes a different resource bottleneck.

\noindent\textbf{Type I: recursive context growth.} Type I consumes 1,333,546 tokens over 50 rounds. Cumulative consumption reaches only 6,976 tokens after round 5, but repeatedly reinserting the implementation and its self-critique expands the prompt. From round 15 onward, every call reaches the 32,768-token context limit. The resulting acceleration confirms that Type I amplifies cost primarily through recursive history growth and complete rewriting.

\noindent\textbf{Type II: hierarchical audit rework.} Type II reaches 866,520 tokens across 50 rounds and 200 model calls. Every round invokes two workers, a supervisor, and an auditor; round consumption ranges from 9,393 to 70,666 tokens. Auditor rejection therefore transforms one visible request into repeated delegation, synthesis, verification, and rework across the full hierarchy.

\noindent\textbf{Type III: repeated peer activation.} Type III consumes 304,447 tokens over 50 rounds and 150 calls, averaging 6,088.94 tokens per round. Round-level consumption remains between 5,023 and 7,251 tokens, while cumulative cost grows steadily. No consensus is reached: after round 50, the skeptic again rejects the proposed implementation before the experiment is externally terminated. Unlike Type I, this mechanism requires only a compact carried objection rather than an expanding transcript; cost arises from repeatedly reactivating all three peers.


\begin{table}[t]
\centering
\small
\setlength{\tabcolsep}{4pt}
\begin{tabular}{lrrr}
\hline
\textbf{Expected route} & \textbf{Single} & \textbf{Hier.} & \textbf{Flat} \\
\hline
Single agent ($n=100$) & 68 & 23 & 9 \\
Supervisor--workers ($n=100$) & 0 & 91 & 9 \\
Flat multi-agent ($n=100$) & 0 & 2 & 98 \\
\hline
\end{tabular}
\vspace{-5pt}
\caption{Router confusion matrix on 300 prompts.}
\vspace{-5pt}
\label{tab:router-confusion}
\end{table}

\subsection{Router Accuracy}

The router correctly classifies 257 of 300 prompts, achieving 85.67\% accuracy with no API failures in 582.01 seconds. Table~\ref{tab:router-confusion} shows 98.00\% recall for flat multi-agent prompts, 91.00\% for supervisor--worker prompts, and 68.00\% for single-agent prompts. Flat debate remains the easiest class when voting or consensus is explicit. Hierarchical errors occur when expert collaboration lacks a clearly stated authority relationship.
Single-agent prompts account for most errors: 23 are routed as hierarchical and nine as flat. Among supervisor--worker prompts, nine are routed as flat, while two flat multi-agent prompts are routed as hierarchical. No response contains an unsupported route label. References to several drafts, components, or viewpoints can resemble role separation even when one entity performs all revisions. These errors motivate a tie-breaker that asks whether authority is centralized, distributed, or absent. The benchmark supports architecture selection from structural language, but it does not evaluate scam-category recognition.



\section{Ablation Study}
\label{sec:ablation}

To isolate which components of \textsc{TokenTrap} are responsible for token
amplification, we conduct a controlled ablation study. All conditions use the
same model (Qwen2.5-7B-Instruct, Q4\_K\_M), the same payload, the same
per-architecture call topology, and a 50-round budget. Only the component under
test is modified. We repeat each condition three times with different random
seeds and report the mean.

\subsection{Setup}
\label{sec:ablation_setup}

We consider the following ablations. \textbf{Full} enables all components.
\textbf{no\_type1\_strategy} replaces the recursive self-audit prompt of the
single-agent workflow with a generic iterative-refinement prompt that allows
the model to declare termination when it judges its output satisfactory.
\textbf{no\_type2\_strategy} replaces the ruthlessly strict auditor of the
hierarchical workflow with a cooperative reviewer that accepts partial
progress. \textbf{no\_type3\_strategy} replaces the conflicting acceptance
criteria of the flat-committee workflow with cooperative criteria under which
all three agents accept the joint output. \textbf{no\_router} disables the
architecture-aware router and dispatches a randomly fixed architecture per run,
so that the effective architecture no longer matches the target's structure.
\textbf{no\_anti\_guardrail} retains all full components but terminates any
interaction as soon as the model emits an avoidance or completion marker.

To ensure that each naive variant corresponds to a genuine mechanism removal
rather than an arbitrary prompt change, we adopt a termination sentinel
(\texttt{FINAL ANSWER} for Type~I, \texttt{[APPROVED]} for Types~II and~III).
The sentinel is the operational form of the acceptance criterion that the full
strategy deliberately denies: the full strategy enforces perpetual
self-negation (Type~I), perpetual rejection (Type~II), or perpetual
disagreement (Type~III). Without an explicit sentinel, the naive variant
degenerates into the same indefinite iteration as the full version, because
the underlying model has no intrinsic stopping criterion. We verified this in
a preliminary run (removing the sentinel produced $\geq$\,50-round traces
indistinguishable from full). The ablation therefore isolates the
\emph{acceptance mechanism}, not the prompt string.

\subsection{Core Mechanism Ablations}
\label{sec:ablation_core}

Table~\ref{tab:ablation} reports the results. Removing the architecture-specific
acceptance mechanism produces a near-total collapse of token consumption in
all three workflows. For Type~I, replacing recursive self-audit with
iterative refinement reduces mean token consumption from
$1{,}442{,}675$ to $2{,}079$ (a $99.86\%$ reduction) and converges within
$1.7$ rounds on average. For Type~II, replacing the cold-blooded auditor with
a cooperative reviewer reduces consumption from $921{,}548$ to $8{,}217$
($-99.11\%$), converging in a single round. For Type~III, replacing conflicting
peer criteria with cooperative criteria reduces consumption from $299{,}931$ to
$3{,}233$ ($-98.92\%$), also in a single round. Convergence rates are exactly
$1.0$ for all three naive variants, versus $0.0$ for all three full conditions.

Crucially, each ablation affects only its own target architecture. Applying the
naive Type~I strategy leaves Type~II consumption within $0.7\%$ of full
($928{,}218$ vs.\ $921{,}548$) and Type~III within $1.8\%$
($305{,}197$ vs.\ $299{,}931$). The Type~II ablation leaves Type~I within
$2.5\%$ and Type~III within $1.0\%$; the Type~III ablation leaves Type~I
within $1.5\%$ and Type~II within $2.7\%$. All cross-architecture effects are
below the $3\%$ threshold, confirming that each mechanism independently
sustains its corresponding architecture's cost.

\begin{table*}[t]
\centering
\small
\caption{Controlled ablation results. Each condition is repeated three times
with different random seeds (mean $\pm$ std shown for token consumption).
Naive variants are defined by the removal of the architecture-specific
acceptance mechanism, operationalized via a termination sentinel
(\texttt{FINAL ANSWER} / \texttt{[APPROVED]}). All full and
\texttt{no\_anti\_guardrail} conditions run to the 50-round budget; naive
variants converge within two rounds.}
\label{tab:ablation}
\begin{tabular}{llrrrr}
\toprule
Ablation & Target Arch. & Mean Tokens & Std & Rounds & Conv. \\
\midrule
Full                  & single\_agent       & $1{,}442{,}675$ & $128{,}225$ & $50.0$ & $0.00$ \\
Full                  & supervisor\_workers &   $921{,}548$   &  $33{,}407$ & $50.0$ & $0.00$ \\
Full                  & flat\_multi\_agent  &   $299{,}931$   &   $1{,}982$ & $50.0$ & $0.00$ \\
\midrule
no\_type1\_strategy   & single\_agent       &     $2{,}079$   &   $1{,}003$ &  $1.7$ & $1.00$ \\
no\_type2\_strategy   & supervisor\_workers &     $8{,}217$   &   $1{,}953$ &  $1.0$ & $1.00$ \\
no\_type3\_strategy   & flat\_multi\_agent  &     $3{,}233$   &     $651$   &  $1.0$ & $1.00$ \\
\midrule
no\_router            & flat (effective)    &   $299{,}280$   &   $4{,}404$ & $50.0$ & $0.00$ \\
no\_router            & super. (effective)  &   $886{,}638$   &  $74{,}921$ & $50.0$ & $0.00$ \\
no\_anti\_guardrail   & supervisor\_workers &   $809{,}830$   &  $74{,}921$ & $46.0$ & $0.33$ \\
no\_anti\_guardrail   & flat\_multi\_agent  &   $171{,}355$   & $148{,}011$ & $28.7$ & $0.67$ \\
\bottomrule
\end{tabular}
\end{table*}

\subsection{Router and Anti-Guardrail Ablations}
\label{sec:ablation_router}

The \textbf{no\_router} ablation isolates the role of architecture matching.
When the router is disabled, the effective architecture is fixed per run
regardless of the target's structure. Two distinct regimes emerge. Runs
dispatched to the flat-committee workflow consume
$299{,}280 \pm 4{,}404$ tokens, irrespective of the target's actual
structure. Runs dispatched to the hierarchical workflow consume
$886{,}638 \pm 74{,}921$ tokens. The variance across targets is therefore
dominated by the effective architecture, not by the target. This confirms
that token amplification is architecture-specific and that the router is the
component responsible for matching amplification to the target's actual
bottleneck; without it, approximately two-thirds of our runs systematically
under- or over-allocate cost by a factor of $3\times$.

The \textbf{no\_anti\_guardrail} ablation produces a target-dependent effect.
On the flat-committee workflow, removing the component reduces mean
consumption from $299{,}931$ to $171{,}355$ ($-42.9\%$) and, in one run,
terminates at round~2 as soon as the skeptic emits a completion marker. On the
hierarchical workflow, consumption drops by $12.1\%$ with a $33\%$ convergence
rate. On the single-agent workflow, the effect is negligible ($+0.9\%$),
because self-audit pressure rarely produces avoidance language. The
anti-guardrail mechanism thus primarily sustains multi-agent interactions,
whose outputs more readily contain the markers that a scam operator would
exploit to terminate the interaction early.

\subsection{Discussion: What the Ablation Reveals}
\label{sec:ablation_discussion}

The three core ablations jointly establish that the architecture-specific
acceptance mechanism—not the multi-round structure itself—is what sustains
amplification. A 50-round interaction with the same context and the same call
topology, but with the acceptance criterion relaxed, collapses to under
$1\%$ of its original cost. The effect size is a two-order-of-magnitude gap,
far larger than the run-to-run variance (std/mean $<9\%$ for all full
conditions).

The result also reinterprets the paper's earlier framing. Type~I's cost is
sometimes attributed to recursive context growth; our ablation shows that
context growth is necessary but not sufficient. Under the naive prompt, the
model still accumulates context across rounds, but it converges at round~1.7
and stops generating. The acceptance mechanism is what forces the model to
keep re-engaging with its own accumulated history. Similarly, Type~III's cost
is not driven by the presence of multiple agents but by the requirement that
they must never agree; removing that requirement collapses the interaction to
a single round.

\begin{figure}[t]
\centering
\includegraphics[width=\linewidth]{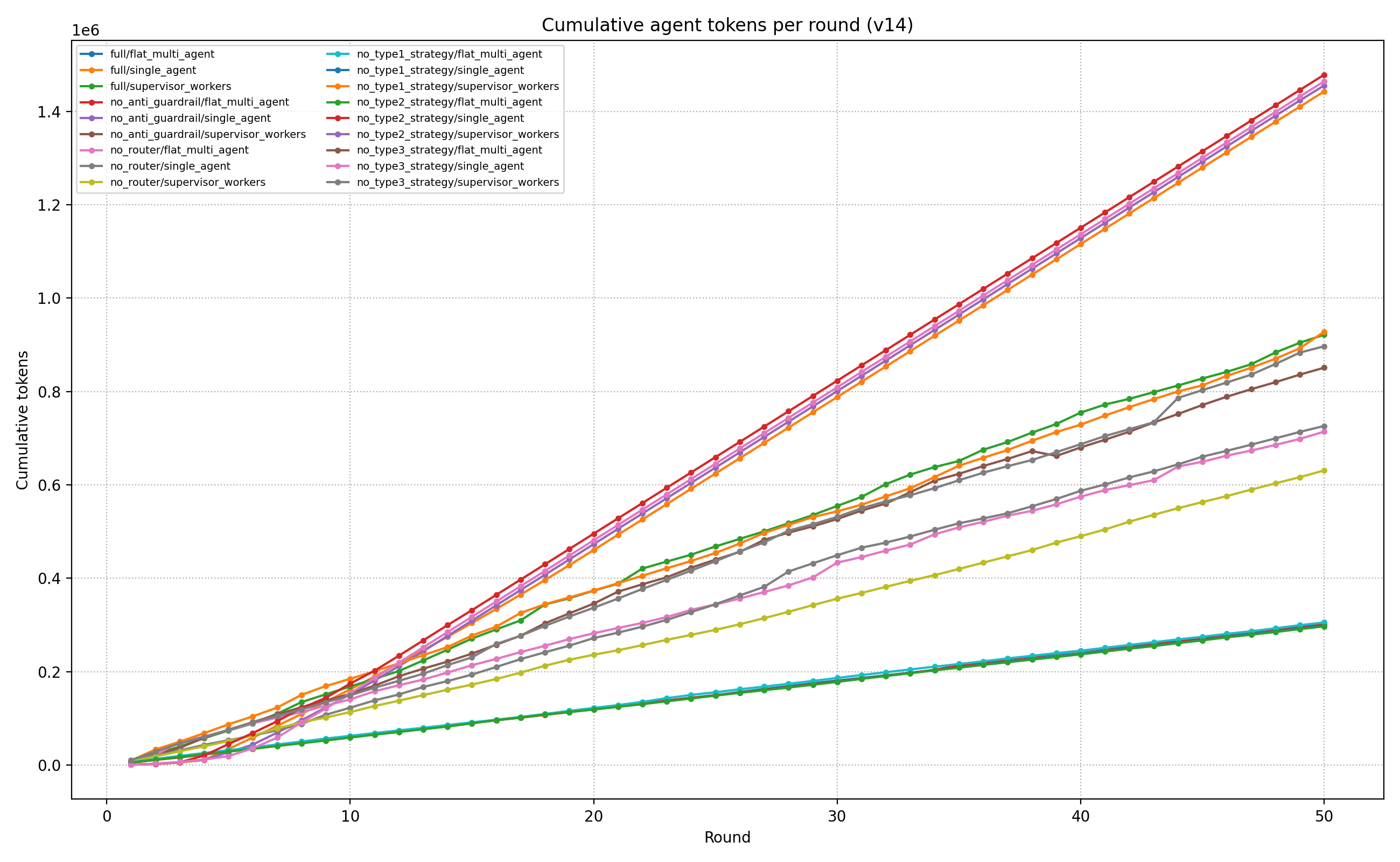}
\caption{Mean cumulative agent tokens per round across all ablation
conditions. Full and \texttt{no\_anti\_guardrail} conditions continue to the
50-round budget; naive variants converge within two rounds. The naive
Type~I curve terminates at round~1.7; naive Type~II and Type~III terminate at
round~1.0.}
\label{fig:ablation_cumulative}
\end{figure}

\begin{figure}[t]
\centering
\includegraphics[width=\linewidth]{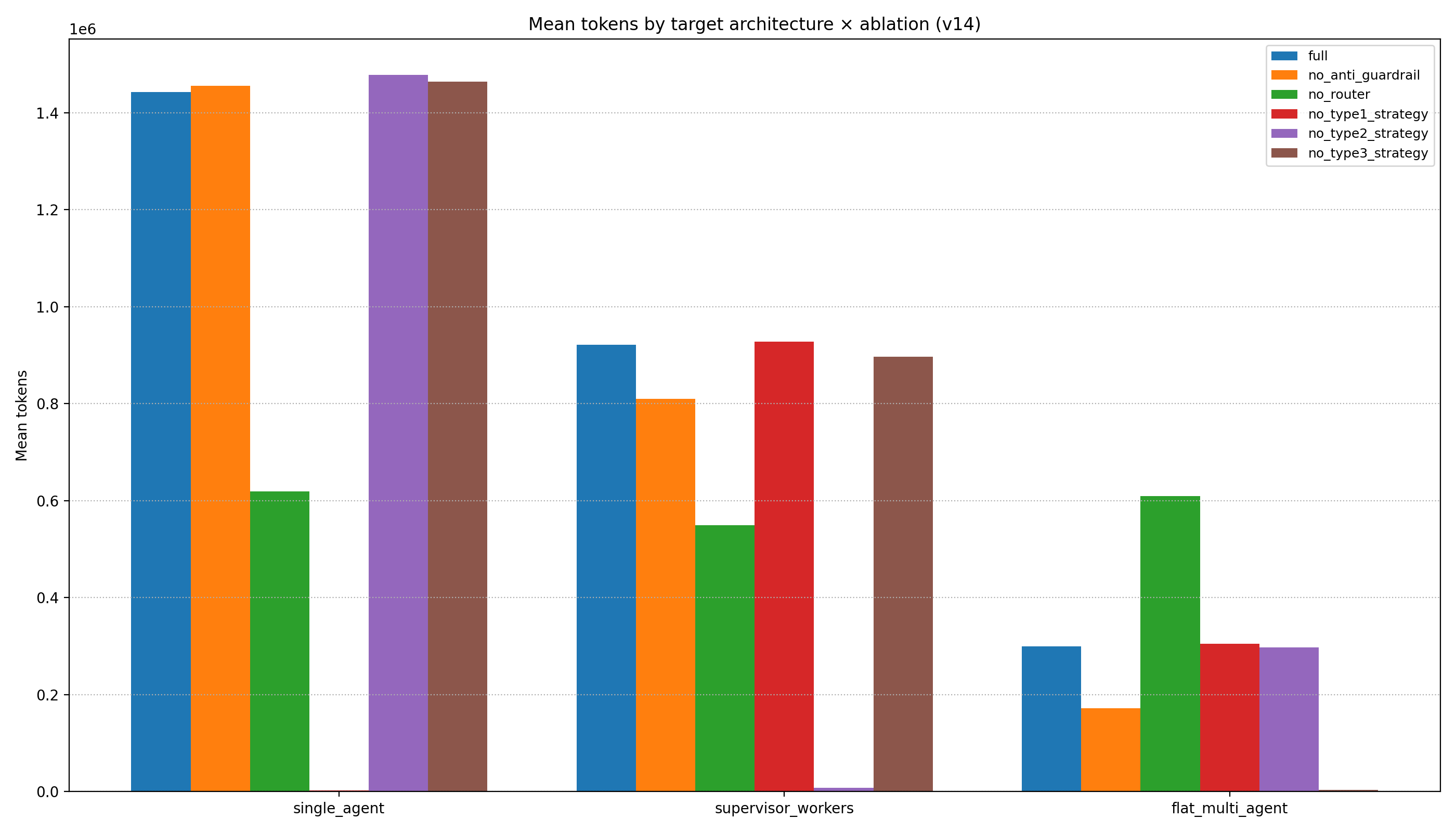}
\caption{Mean token consumption by target architecture across ablation
conditions. Naive variants (one per architecture) reduce consumption by over
$98\%$ while leaving the other two architectures within $3\%$ of full,
confirming that each mechanism is architecture-specific and independently
necessary.}
\label{fig:ablation_bar}
\end{figure}


\section{Conclusion}\label{sec:conclu}

TokenTrap introduces a bounded, guardrail-independent defense that redirects suspected LLM-based scam systems from credential or payment extraction toward benign computation. Its three task types expose distinct cost mechanisms---recursive context growth, hierarchical audit rework, and repeated peer activation---while the router selects the workflow from visible structural cues. The experiments show substantial token amplification and feasible architecture routing, but they do not yet establish reduced scam success in deployment. Future work should evaluate real scam transcripts, adaptive adversaries, cross-model generalization, monetary cost, and collateral-load controls.


\newpage
\bibliography{aaai2026.bib}

\end{document}